\documentclass[fleqn,usenatbib]{mnras}

\usepackage{newtxtext,newtxmath}

\usepackage[T1]{fontenc}

\DeclareRobustCommand{\VAN}[3]{#2}
\let\VANthebibliography\thebibliography
\def\thebibliography{\DeclareRobustCommand{\VAN}[3]{##3}\VANthebibliography}

\usepackage{graphicx}	
\usepackage{amsmath}	
\usepackage{tabularx}

\usepackage{textgreek}
\usepackage{multirow}

\title[SN 2023tsz: A SN Ibn in a very low-mass host]{SN 2023tsz: A helium-interaction driven supernova in a very low-mass galaxy
}

\author[Warwick et al.]{
B. Warwick,$^{1}$\thanks{E-mail: ben.warwick@warwick.ac.uk}
J. Lyman,$^{1}$
M. Pursiainen,$^{1}$ 
D. L. Coppejans,$^{1}$
L. Galbany,$^{2, 3}$
G. T. Jones,$^{1}$
T. L. Killestein,$^{4, 1}$
\newauthor
A. Kumar,$^{1}$
S. R. Oates,$^{5}$
K. Ackley,$^{1}$
J. P. Anderson,$^{6, 7}$
A. Aryan,$^{8}$
R. P. Breton,$^{9}$
T. W. Chen,$^{8}$
P. Clark,$^{10}$
\newauthor
V. S. Dhillon,$^{11, 12}$
M. J. Dyer,$^{11}$
A. Gal-Yam,$^{13}$
D. K. Galloway,$^{14, 15}$
C. P. Guti{\'{e}}rrez,$^{3, 2}$
M. Gromadzki,$^{16}$
\newauthor
C. Inserra,$^{17}$
F. Jim{\'{e}}nez-Ibarra,$^{14}$
L. Kelsey,$^{10}$
R. Kotak,$^{4}$
T. Kravtsov,$^{4}$
H. Kuncarayakti,$^{4, 18}$
M. R. Magee,$^{1}$
\newauthor
K. Matilainen,$^{4}$
S. Mattila,$^{4, 19}$
T. E. M{\"{u}}ller-Bravo,$^{2, 3}$
M. Nicholl,$^{20}$
K. Noysena,$^{21}$
L. K. Nuttall,$^{10}$
P. O'Brien,$^{22}$
\newauthor
D. O'Neill,$^{1}$
E. Pall{\'{e}},$^{12}$
T. Pessi,$^{6}$
T. Petrushevska,$^{23}$
G. Pignata,$^{24}$
D. Pollacco,$^{1}$
F. Ragosta,$^{25, 26}$
G. Ramsay,$^{27}$
\newauthor
A. Sahu,$^{1}$
D. K. Sahu,$^{28}$
A. Singh,$^{29, 30}$
J. Sollerman,$^{30}$
E. Stanway,$^{1}$
R. Starling,$^{22}$
D. Steeghs,$^{1}$
R. S. Teja,$^{28, 31}$
\newauthor
K. Ulaczyk$^{1}$
\\
Affiliations are listed at the end of the paper.
}

\date{Accepted XXX. Received YYY; in original form ZZZ}

\pubyear{2024}

\begin{document}
\label{firstpage}
\pagerange{\pageref{firstpage}--\pageref{lastpage}}
\maketitle

\begin{abstract}
SN\,2023tsz is a Type Ibn supernova (SNe Ibn) discovered in an extremely low-mass host. SNe Ibn are an uncommon subtype of stripped-envelope core-collapse SNe. They are characterised by narrow helium emission lines in their spectra and are believed to originate from the collapse of massive Wolf-Rayet (WR) stars, though their progenitor systems still remain poorly understood. In terms of energetics and spectrophotometric evolution, SN\,2023tsz is largely a typical example of the class, although line profile asymmetries in the nebular phase are seen, which may indicate the presence of dust formation or unshocked circumstellar material. Intriguingly, SN\,2023tsz is located in an extraordinarily low-mass host galaxy that is in the 2nd percentile for SESN host masses and star formation rates (SFR). The host has a radius of 1.0\,kpc, a $g$-band absolute magnitude of $-12.73$, and an estimated metallicity of $\log(Z_{*}/Z_{\odot}$) = $-1.56$. The SFR and metallicity of the host galaxy raise questions about the progenitor of SN\,2023tsz. The low SFR suggests that a star with sufficient mass to evolve into a WR would be uncommon in this galaxy. Further, the very low-metallicity is a challenge for single stellar evolution to enable H and He stripping of the progenitor and produce a SN Ibn explosion. The host galaxy of SN\,2023tsz adds another piece to the ongoing puzzle of SNe Ibn progenitors, and demonstrates that they can occur in hosts too faint to be observed in contemporary sky surveys at a more typical SN\,Ibn redshift.

\end{abstract}

\begin{keywords}
(stars:) supernovae: general -- (stars:) circumstellar matter -- stars: massive
\end{keywords}



\section{Introduction} \label{Intro}

Type Ibn supernovae (SNe Ibn) are a subclass of supernovae (SNe) that are characterised by the presence of narrow helium (He) emission lines in their spectra but the absence of strong hydrogen (H) features  \citep[e.g.][]{2017hsn..book..403S, 2017hsn..book..195G}. These spectral properties are explained by the interaction of supernova ejecta with He-rich, but H-poor, circumstellar material (CSM). The first SN Ibn was observed in 1999 \citep[SN 1999cq,][]{2000AJ....119.2303M}. However, the label was not coined until the analysis of SN 2006jc \citep{2007ApJ...657L.105F, 2007Natur.447..829P}, which is considered the prototypical SN Ibn. Despite nearly two decades since their identification, there are still questions about the nature of their progenitors. These questions persist due to the lack of a confirmed direct detection of a SN Ibn progenitor as this type is relatively rare, with only 66 classified to date.\footnote{Based on a Transient Name Server, https://www.wis-tns.org/, query on 18/06/2024} It was estimated by \cite{2022ApJ...927...25M} that they make up around 1\% of core-collapse (CC) SNe, with an observed rate of $~3\%$ \citep{2020ApJ...904...35P}. 

The original progenitor suggested for the prototypical SN Ibn, SN 2006jc was a H-poor massive Wolf-Rayet (WR) star embedded in a He-rich CSM \citep{2007Natur.447..829P, 2008ApJ...687.1208T}. Such progenitors are proposed for SNe Ibn, as the mass loss that WR stars undergo prior to explosion can explain the properties seen in SN Ibn light curves \citep{2022ApJ...927...25M}. This progenitor model is supported by the fact that the majority of SN Ibn are found in active star forming regions \citep[e.g.,][]{2015A&A...580A.131T, 2015MNRAS.449.1941P}. However, there is a notable exception, PS1-12sk, which was discovered in the outskirts of an elliptical galaxy in a region with a low star formation rate (SFR) \citep{2013ApJ...769...39S, 2019ApJ...871L...9H}. Another proposed progenitor for SNe Ibn is a low-mass, $\lesssim5\,M_{\odot}$, He star likely arising from a binary system. In this scenario, binary interactions are the cause of the mass loss that creates the CSM prior to explosion. The Type Ibn supernova then arises from either the core collapse of the He star, or an explosion triggered by the merger of the He star and its binary companion \citep{2016ApJ...833..128M, 2022A&A...658A.130D, dong2024sn2023fyq}. Such a progenitor could be more likely in an older stellar population. It is also plausible that multiple progenitor channels could lead to SNe that would be classified as a SN Ibn, such as the confirmed case of a SN IIb that exploded inside a dense CSM and appeared to be a SN Ibn \citep{2020MNRAS.499.1450P}. Consequently, the nature and homogeneity of SN Ibn progenitors remains uncertain.

Compared to most CC SNe, which are powered primarily through the decay of $^{56}$Ni produced during the explosion, a combined nickel decay and CSM interaction model is required to explain the lightcurves of SN Ibn \citep{2020MNRAS.492.2208C}. The CSM interaction is dominant at early times of SNe Ibn evolution, and is thought to explain their shorter time-scales for both the rise and decline of their lightcurves. These shorter time-scales make SN Ibn observationally rare as they are harder to characterise upon discovery, and are particularly hard to capture on their rise to peak. The current generation of all-sky surveys, which are observing at higher cadences than previous surveys, will help to find and characterise SNe Ibn earlier in their evolution allowing improved probing of their physics.

\begin{figure}
    \centering
    \includegraphics[width=\columnwidth]{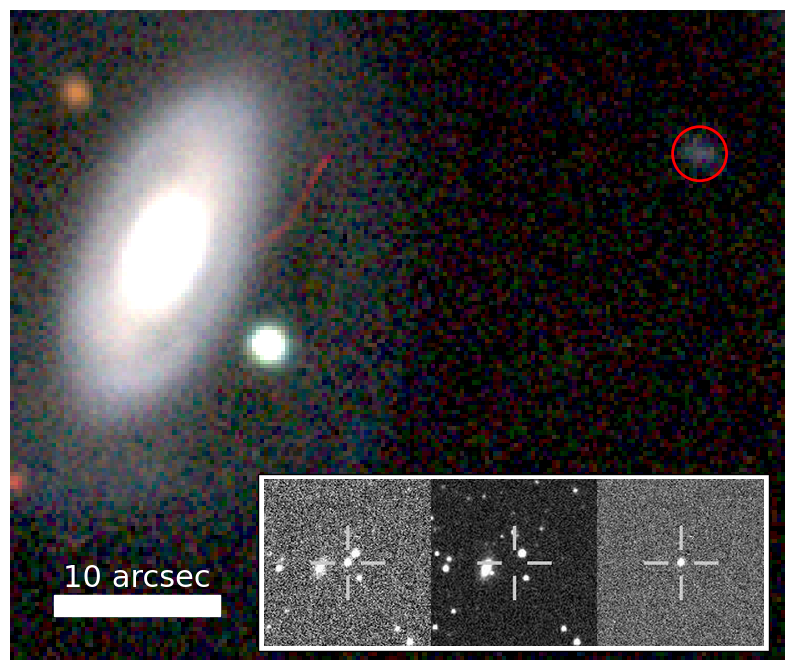}
    \caption{Image showing the host of SN 2023tsz (circled in red) and its immediate surroundings, including LEDA 152972. The image was created using $g$, $r$, and $i$ observations from DR9 of the DESI Legacy Survey \citep{2019AJ....157..168D}. The red circle is the radius around SN 2023tsz's host that contains 99\% of its light in the $g$ band. The three image cutouts in the bottom right of the image show the science (\textit{left}), template (\textit{middle}), and difference (\textit{right}) images from GOTO that were used to discover SN 2023tsz. The GOTO science image was taken $+3.2$\,d with respect to our estimate of the peak (see text).}
    \label{fig:tsz_loc}
\end{figure}

The focus of this paper is the Type Ibn supernova SN 2023tsz. It was discovered by the Gravitational-wave Optical Transient Observer \citep[GOTO;][]{2022MNRAS.511.2405S, 2022SPIE12182E..1YD, 2024arXiv240717176D} on 2023 September 28 and reported to the Transient Name Server under the name GOTO23anx \citep{2023TNSTR2419....1G}. There was a prior non-detection and two detections from All-sky Automated Survey for Supernovae \citep[ASAS-SN;][]{2014ApJ...788...48S, 2017PASP..129j4502K} on September 13, 19, and 25 respectively. The SN is associated with a faint galaxy, a possible satellite of LEDA 152972, located $33^{\prime\prime}$ from its centre, visible in the DESI Legacy DR9 images \citep{2019AJ....157..168D} as shown in Figure \ref{fig:tsz_loc}. It is also the first object to be discovered, reported \citep{2023TNSTR2419....1G}, and classified \citep{2023TNSCR2442....1P} entirely by the GOTO collaboration. 

This paper is structured as follows: Section \ref{Obs} presents the observations and data reduction methods, Section \ref{Analysis} presents the analysis of the data, Section \ref{Disc} discusses the implications of our analysis, and Section \ref{Con} concludes our findings. Throughout this paper we have corrected for a Milky Way extinction of E$_{B-V}$ = $0.036$\,mag \citep{2011ApJ...737..103S}. When analysing the spectra in Section \ref{Spec Analysis} there is no evidence for significant host extinction. We assume a flat $\Lambda$CDM cosmology with $\Omega_{m}$ = $0.31$ and H$_{0}$ = 67.66 km s$^{-1}$ Mpc$^{-1}$ \citep{2020A&A...641A...6P}. In the absence of host lines we adopt the redshift value of z = 0.028, derived from spectrum template matching using the Supernova Identification code \citep[SNID;][]{2011ascl.soft07001B}, see Section \ref{Spec Analysis}. Using this cosmology and redshift we calculate a luminosity distance for SN\,2023tsz of $130$\,Mpc.

\section{Observations and Data Reduction} \label{Obs}

In addition to GOTO $L$ band (described in \citealt{2022MNRAS.511.2405S}), photometry of SN 2023tsz was collected in bands $ugriz$ using the Liverpool Telescope \citep[LT;][]{2004SPIE.5489..679S}, the Rapid Eye Mount telescope \citep[REM;][]{2004SPIE.5492.1613C}, and Las Cumbres Observatory \citep[LCO;][]{2013PASP..125.1031B} Global Telescope Network. The data from LT and LCO were provided already pre-reduced for bias, dark and flat-field corrections using their own pipelines \citep{curtis_mccully_2018_1257560}. The data from REM were reduced using our own pipeline. The light curves from these observations were calculated using the photometry-sans-frustration pipeline \citep[\texttt{psf};][]{2023ApJ...954L..28N} making use of the inbuilt template subtraction of \texttt{psf}. Further photometry was collected in the near UV bands $uvm2$, $uvw2$, $uvw1$ with the Neil Gehrels Swift Observatory (\textit{Swift}) UV-Optical telescope \citep[UVOT;][]{2005SSRv..120...95R}, $g$ from ASAS-SN, and $o$ from the Asteroid Terrestrial-impact Last Alert System \citep[ATLAS;][]{2018PASP..130f4505T, 2020PASP..132h5002S}. The ASAS-SN light curve was generated using their Sky Patrol service \footnote{https://asas-sn.osu.edu/}. The ATLAS light curve was generated using their forced photometry server \citep{2021TNSAN...7....1S}. The light curves from \textit{Swift} UVOT were reduced using a $7^{\prime\prime}$ aperture to extract the photometry. This aperture was used due to a slight issue with \textit{Swift} that caused smearing of the sources on the images \citep{2023GCN.34633....1C}.

Spectra of SN 2023tsz were obtained using the Alhambra Faint Object Spectrograph and Camera (ALFOSC) on the Nordic Optical Telescope (NOT), the Himalayan Faint Object Spectrograph Camera (HFOSC) on the Himalayan Chandra Telescope (HCT), the ESO Faint Object Spectrograph Camera 2 (EFOSC2) on the New Technology Telescope (NTT) at La Silla Observatory, taken by the extended Public ESO Spectroscopic Survey for Transient Objects plus \citep[ePESSTO+;][]{2015A&A...579A..40S}, and the Optical System for Imaging and low-Intermediate-Resolution Integrated Spectroscopy plus (OSIRIS+) on the Gran Telescopio Canarias (GTC). The data from ALFOSC were reduced using the \texttt{PyNOT-redux} reduction pipeline\footnote{github.com/jkrogager/PyNOT/}. The spectroscopic data from HFOSC were reduced in a standard manner using the packages and tasks in \texttt{IRAF} with the aid of the Python scripts hosted at \textsc{RedPipe} \citep{2021redpipe}. The data from EFOSC2 was reduced using the PESSTO pipeline \footnote{https://github.com/svalenti/pessto} \citep{2015A&A...579A..40S}. GTC optical spectra from OSIRIS+ was reduced following the routines in Piscarrera et al. (in prep.). based on PypeIt \citep{2020JOSS....5.2308P}. The spectral log is provided in Table \ref{tab:Spec_log}.

Photometry of the host of SN\,2023tsz was obtained from the Kilo-Degree Survey \citep[KiDS;][]{2015A&A...582A..62D}, in the $u$, $g$, $r$ and $i$ bands, and from the VISTA Kilo-degree Infrared Galaxy survey \citep[VIKING;][]{2013Msngr.154...32E} in the $z$ band, detailed in Table \ref{tab:Surv_log}. We also obtain upper limits from the Galaxy Evolution Explorer \citep[GALEX;][]{2005ApJ...619L...1M} in the FUV and NUV bands, and the Wide-field Infrared Survey Explorer \citep[WISE;][]{2010AJ....140.1868W} in the WISE 1, 2, 3, and 4 bands. Finally, one epoch of ALFOSC V-band imaging polarimetry was obtained. The data were reduced and analysed following \cite{2023A&A...674A..81P}.

\section{Analysis} \label{Analysis}

\subsection{Photometry}

\begin{figure*}
    \centering
    \includegraphics[width=\textwidth]{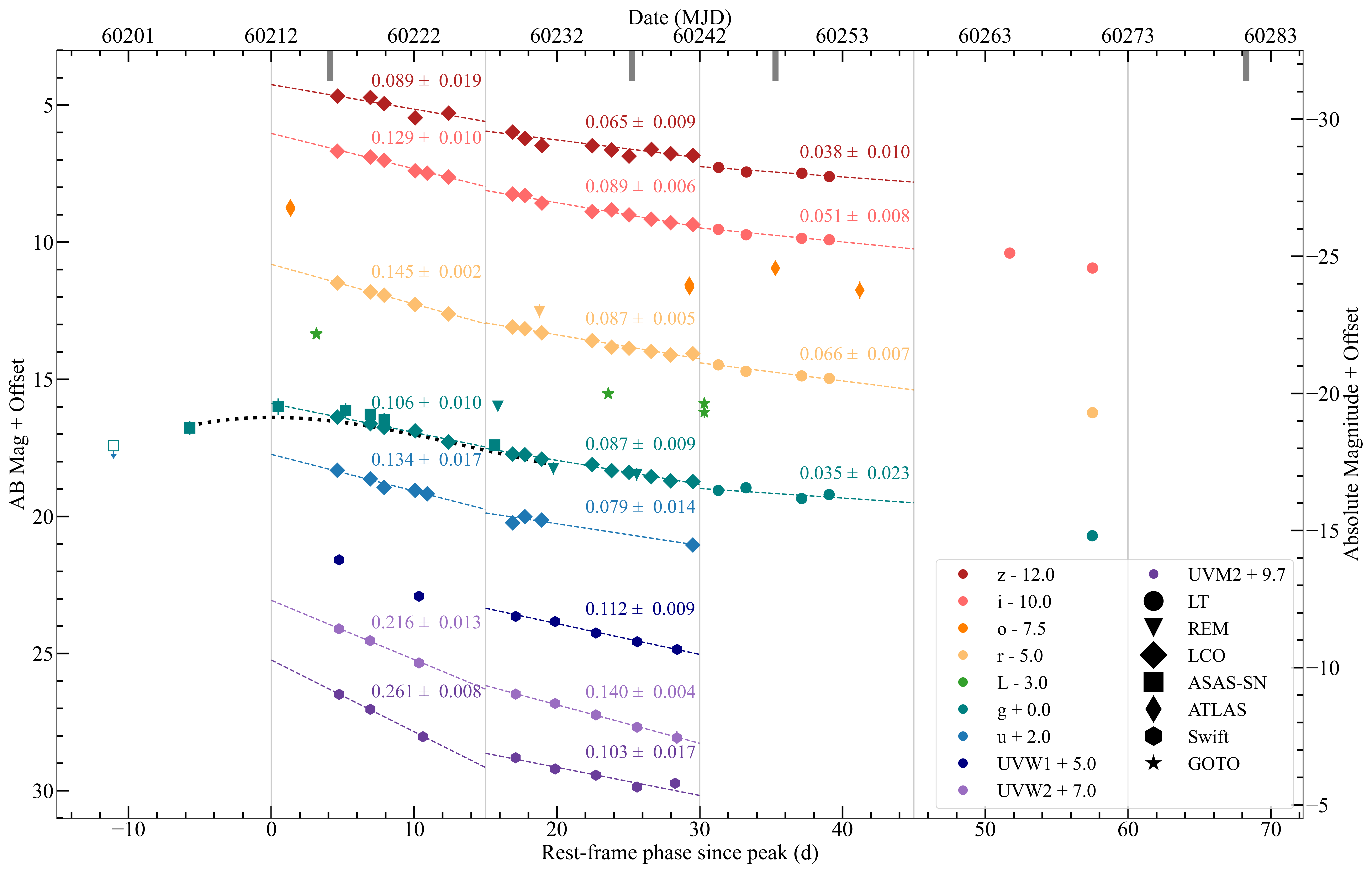}
    \caption{The multi-band light curve of SN 2023tsz. In the legend, the number denotes the offset applied to the values in that filter. The 3rd order polynomial used to fit the peak date in the $g$-band is shown by the black dotted curve. Dashed lines are used to show the light curve decline in each band over 15 day intervals after peak. The decline rate for each interval (in mag d$^{-1}$) is indicated on the plot next to the dashed line. Each 15 day interval is shown by the vertical light grey lines. We only fit the decline if there are at least 3 observations in the 15 day increment. The thick grey lines at the top of the plot represent the epochs when spectroscopic observations were obtained.}
    \label{fig:Multi_band_LC}
\end{figure*}

\begin{figure}
    \centering
    \includegraphics[width=\columnwidth]{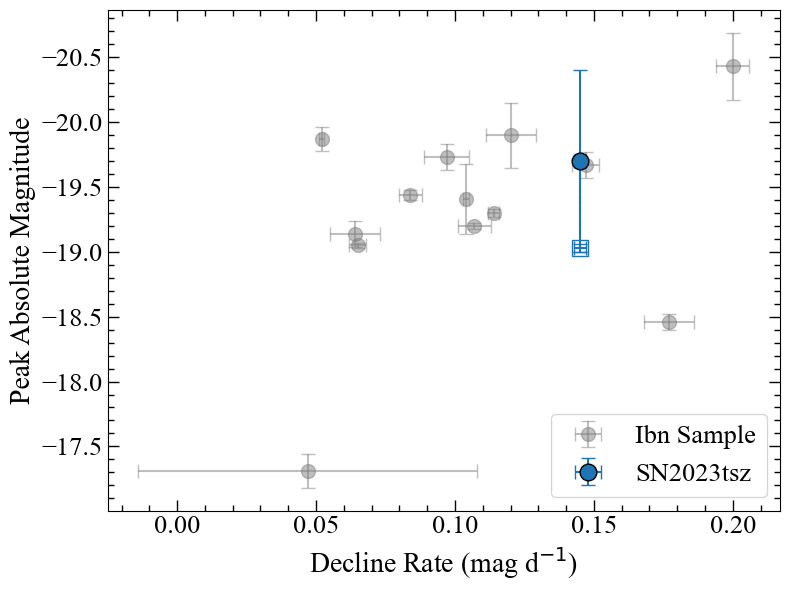}
    \caption{The initial 15 day post peak decline rate and observed peak absolute magnitude of SN\,2023tsz compared with other SNe Ibn. The two markers for SN\,2023tsz encapsulate the uncertainty from not capturing the peak in the r band. The filled marker is an estimate of peak by extrapolating the decline rate back to the epoch of the g-band peak, and the open square symbol is a certain lower limit on the absolute magnitude from the r-band observation at +4.76d. Although we are comparing $r$-band observations to an $R$-band data set, differences between the filter throughputs are not at a level to change our inferences. The data for the SNe Ibn sample is from \protect\cite{2017ApJ...836..158H}. }
    \label{fig:Comparison}
\end{figure}

The multi-band light curve of SN 2023tsz is shown in Figure \ref{fig:Multi_band_LC}. Only the ASAS-SN $g$ band had pre-peak observations, one a non-detection and the other a detection. This is due to the fact that SN 2023tsz was discovered as it exited Solar conjunction, resulting in a late discovery and corresponding poor pre-peak observations. The decline was well sampled in multiple bands up to $\sim 40$ days post peak, with later epochs sampled in $g$, $r$ and $i$. We define our peak epoch as the peak in the $g$ band obtained from fitting a 3rd order polynomial to the data within +/-20 days of the observed peak. This fitting gave us a peak date of MJD $60212\pm5$. The large error on the time of the peak is due to the lack of constraining observations; going forward we adopt MJD $60212$ as our peak date and use this for phase = $0$.  To estimate its decline rate in 15 day increments post peak, we performed a linear regression on the data. These 15 day increments were chosen to investigate the visible flattening of the light curve, and to allow for comparison to other SN Ibn from \cite{2017ApJ...836..158H}.  The results of this are shown in Figure \ref{fig:Multi_band_LC}. In the sample analysis of \cite{2017ApJ...836..158H}, decline rates were estimated mostly in $R$ band, closely matched with our $r$ band observations. Unfortunately, determining the $r$ band peak brightness is difficult given our data coverage. The observed $r$ band peak was at +4.76d when the SN was \(-19.03 \pm 0.03\)\,mag. The $r$ band decline rate in the first 15d post peak is \(0.145 \pm 0.002\)\,mag\,day\(^{-1}\), which if we extrapolate back to the peak makes the estimated peak $r$ band absolute magnitude of SN\,2023tsz \(-19.7 \pm 0.7\)\,mag. This estimate is likely to be at an epoch before the true $r$ band peak as the SN is cooling and so would peak in the $g$-band earlier than the $r$ band. It is also unlikely that the SN declines at this rate immediately from the peak. Therefore we consider this value as an upper limit on the magnitude. We show the range of peak magnitudes and the initial decline rate alongside a sample of other SN Ibn in Figure \ref{fig:Comparison}. The figure shows that SN\,2023tsz is slightly brighter and faster than the majority of the other SN Ibn from \cite{2017ApJ...836..158H}, but is well within the distribution.

\begin{figure}
    \centering
    \includegraphics[width=\columnwidth]{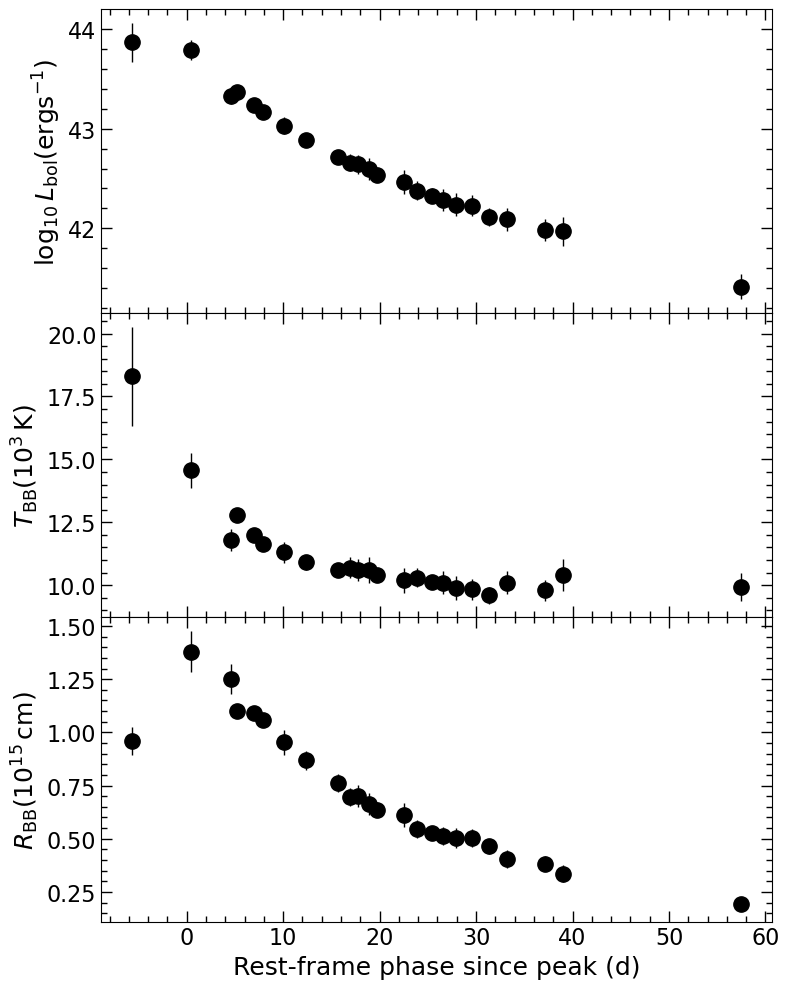}
    \caption{The bolometric light curve (\textit{top}), blackbody temperature (\textit{middle}), and blackbody radius (\textit{bottom}) of SN 2023tsz over all observational epochs.}
    \label{fig:BB_params}
\end{figure}

The bolometric light curve was created using the \textsc{Superbol} routine \citep{2018RNAAS...2..230N}. \textsc{Superbol} works by fitting polynomial light-curve models to each band independently and then interpolating the magnitudes to the times of observation in a defined reference band before using them to fit blackbody models. As the $g$ band has the most observations, and the only pre-peak observations, we used it as our reference band. In bands where we do not have late-time coverage (all bands except $g$, $r$, and $i$), we assume that the colour relative to the $g$ band remains constant at the last measured epoch for that band. The temperatures of the first two epochs, which had only g-band data, were estimated by extrapolating the temperature curve with a second degree polynomial. This was done as at these epochs only $g$ band photometry was obtained resulting in poor temperature estimates from fitting the bolometric light curve. All bands in which observations were obtained, other than the GOTO $L$ band, were used to generate the bolometric light curve. The output of this routine was then smoothed by taking the error-weighted average of the bolometric luminosity (L$_{\mathrm{bol}}$), blackbody temperature (T$_{\mathrm{BB}}$), and blackbody radius (R$_{\mathrm{BB}}$) for each individual day of observations. The full bolometric light curve, along with the corresponding T$_{\mathrm{BB}}$, and R$_{\mathrm{BB}}$ are presented in Figure \ref{fig:BB_params}. 

\begin{figure}
    \centering
    \includegraphics[width=\columnwidth]{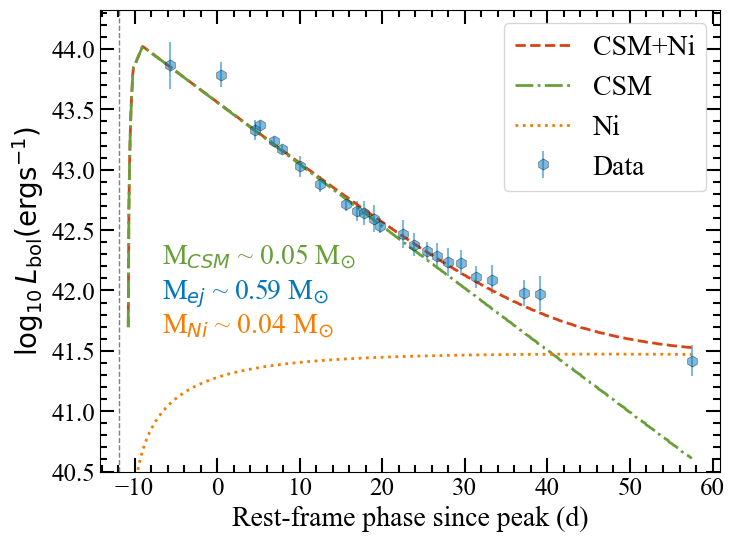}
    \caption{The best-fitting combined CSM+Ni model to the bolometric light curve of SN 2023tsz using the models of \protect\cite{2012ApJ...746..121C}. The individual CSM and Ni models are shown by the green dash-dotted and orange dotted lines respectively. The combined model is shown by the dashed red line. The best fitting CSM (green), ejecta (blue), and Ni (orange) masses are shown on the plot. The vertical dashed line shows the epoch of the ASAS-SN non-detection.}
    \label{fig:BB_fit}
\end{figure}

We fit to the bolometric light curve the semi-analytical models of \cite{2012ApJ...746..121C} for a combined CSM-interaction and $^{56}$Ni decay (CSM+Ni) model, using an \textsc{emcee} bayesian analysis \citep{2013PASP..125..306F}. We set the power-law index for the inner density profile of the ejecta as $\delta=1$, the power-law index for the outer component of the ejecta as $n=10$, the power-law index of the CSM as $s=0$, and the optical opacity of the CSM as $\kappa=0.2$ cm$^{2}$g$^{-1}$. These values were chosen to allow for a  comparison to the work of \cite{2022ApJ...926..125P, 2022ApJ...938...73P}. We use a constant density CSM as, in the model we use, the density profile is degenerate with the radius of the star and changing the density will  effectively only affect the radius. We constrained the explosion epoch to MJD $60203.63 \pm 2.99$. This explosion epoch uses the non-detection, at MJD 60200.64, and first detection, at MJD 60206.62 from ASAS-SN, as its bounds. The best fit model can be seen in Figure \ref{fig:BB_fit} with the best fit model parameters presented in Table \ref{tab:Params_table}. Our second data point lies outside our best fit model, we suspect this is due to ambiguity in the time of the R$_{\mathrm{BB}}$ turn over that occurs around this point. 

\begin{table}
    \renewcommand{\arraystretch}{1.5}
	\centering
	\caption{The best fit parameters for the CSM+Ni model.}
	\label{tab:Params_table}
	\begin{tabular}{lcr} 
		\hline
        Parameter & Description & Value \\
        \hline
        $M_{\mathrm{CSM}}$ ($M_{\odot}$) & CSM mass & $0.05^{+0.05}_{-0.02}$ \\
        $M_{\mathrm{ej}}$ ($M_{\odot}$) & Ejecta mass & $0.59^{+0.69}_{-0.38}$ \\
        E ($10^{51}$ erg) & Total energy of the SN & $0.35^{+0.13}_{-0.12}$ \\ 
        $\dot{M}$ ($10^{-4} M_{\odot}$) & Progenitor mass loss rate & $2.23^{+4.08}_{-1.66}$ \\
        $t_{\mathrm{expl}}$ (d) & Time of explosion relative to peak & $-10.39^{+2.11}_{-1.49}$ \\
        $M_{\mathrm{Ni}}$ ($M_{\odot}$) & Total mass of nickel produced & $0.04^{+0.01}_{-0.01}$ \\
		\hline
	\end{tabular}
\end{table}

\subsection{Spectra} \label{Spec Analysis}

\begin{figure*}
    \centering
    \includegraphics[width=\textwidth]{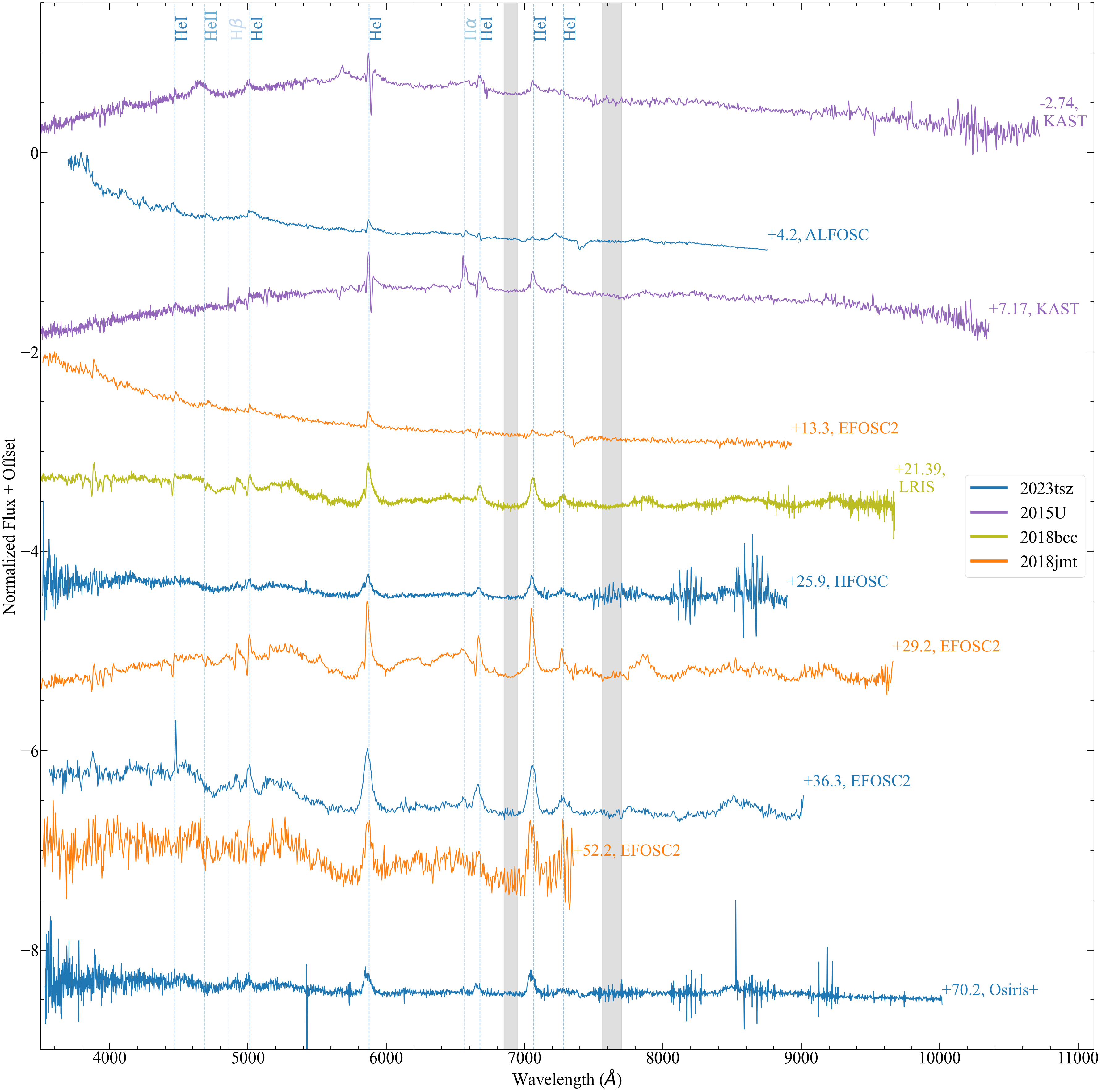}
    \caption{The spectroscopic time series of SN\,2023tsz (blue) compared to the SN Ibn 2015U (purple) \citep{2016MNRAS.461.3057S}, 2018bcc (yellow) \citep{2021A&A...649A.163K}, and 2018jmt (orange) \citep{2018TNSTR1888....1C}. Spectra for these three SNe were obtained from WISeREP \citep{2012PASP..124..668Y}. The phase relative to the peak is shown next to each spectrum. Additionally, the instrument used to obtain each spectrum is listed. The HeI, HeII, H$\alpha$, and H$\beta$ features are marked by dashed lines, and tellurics are shown in grey. The spectra are shown in the rest frame of the supernova.}
    \label{fig:Spectra_Comparison}
\end{figure*}

The spectral time series of SN 2023tsz is compared to example SNe Ibn with spectra taken at similar epochs in Figure \ref{fig:Spectra_Comparison}. SNID \citep{2011ascl.soft07001B} was used to match the initial NOT spectrum of SN\,2023tsz to SNe template spectra and found a best match with a SN Ibn at a redshift of $z$ = 0.028. This template matching, along with manual inspection of the lines present in the spectrum, gave rise to the reported classification based on similarity to several prominent known SN Ibn, also shown in Figure \ref{fig:Spectra_Comparison}. A redshift value of $z$ = 0.028 matches well with a manual inspection of the narrow lines in the spectra. This value differs from the redshift value for LEDA 152972 of $z$ = 0.3515 \citep{2015MNRAS.452.2087L}. This implies a difference of $2200$\,km\,s$^{-1}$ in recession velocity, significantly above the escape velocity expected of LEDA 152972, suggesting a chance alignment with the host of SN\,2023tsz, rather than them being gravitationally bound. However, without narrow host lines we cannot rule this out. The spectra show no presence of suppressed blue emission so we conclude there is no evidence for significant host extinction. 

\begin{figure*}
    \centering
    \includegraphics[width=\textwidth]{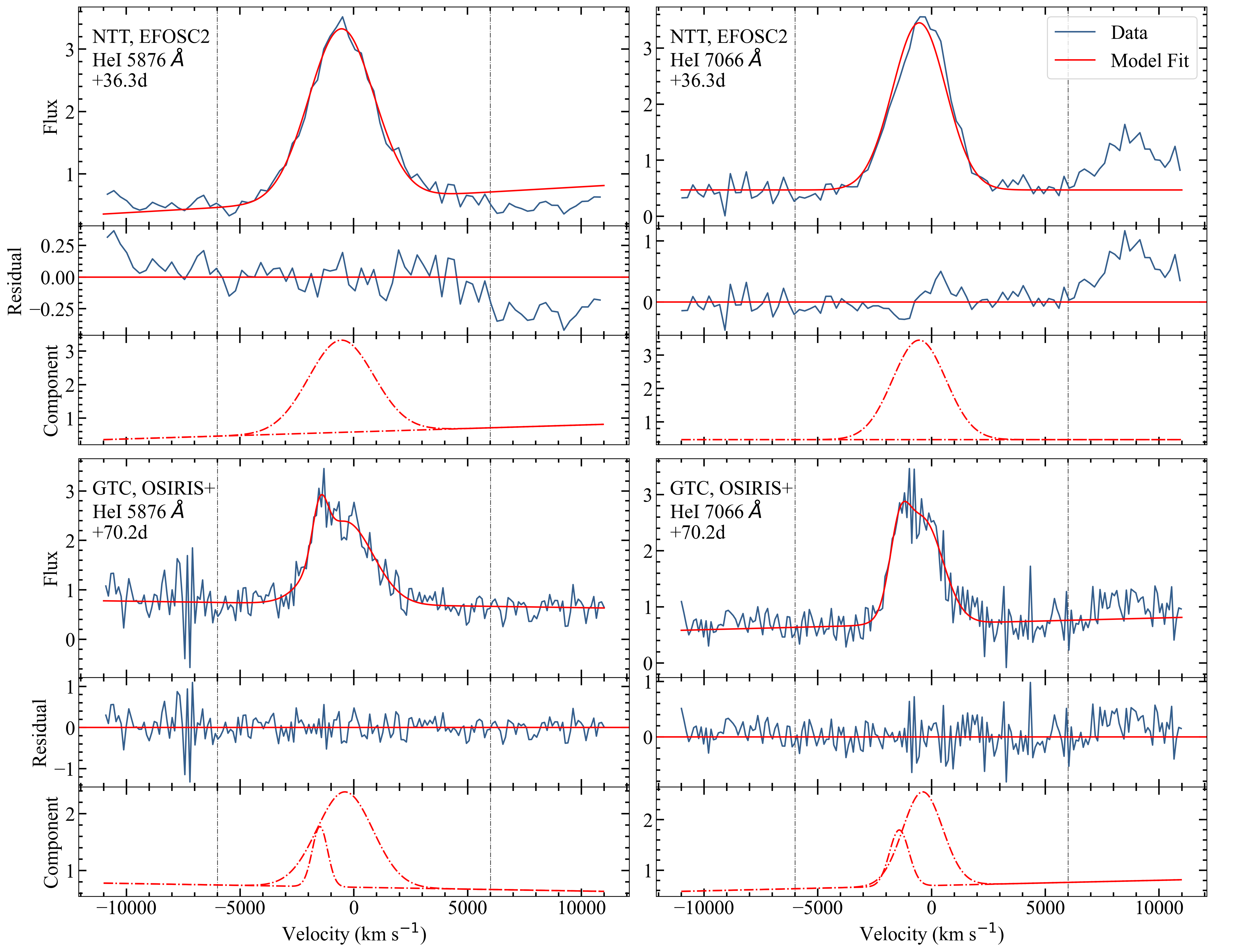}
    \caption{Line fitting of the HeI 5876 \r{A} line (\textit{left}) and HeI 7066 \r{A} line (\textit{right}) for the spectra obtained from the NTT (\textit{top}), and the GTC (\textit{bottom}). Each line is fit with 2 Gaussian profiles, for the NTT spectra the best fit was found to be with a single Gaussian. For each individual plot the top panel shows the combined profile over the spectral feature, the middle panel shows the residuals, and the bottom panel shows the two individual Gaussian components.} 
    \label{fig:Line_fits}
\end{figure*}

The most prominent spectral features across all the spectra are the HeI 5876, 6678, and 7066\,{\AA} lines. To investigate these lines, we fit a Gaussian emission and absorption component, along with an additional Lorentzian emission component, to each line. This fitting was performed over a range extending 6000\,km\,s$^{-1}$ on each side of the line center. The best fit values for each of these components were then determined using an \textsc{emcee} bayesian analysis \citep{2013PASP..125..306F}. Additionally, for the HeI 5876 and 7066{\,\AA} lines in the $+35.3$d and $+70.2$d spectra, we performed the same analysis with two Gaussian emission components (shown in Figure \ref{fig:Line_fits}) to investigate the noticeable blue asymmetry of the spectral features, likely due to a suppression of the red wing of the feature (in the $+70.2$d spectrum). In the classification spectrum, the most prominent line is the He\,I 5876\,{\AA}. In this line we observe a significant P-Cygni profile, with the profile minimum blueshifted by $-1140^{+100}_{-180}\times10^{3}$\,km\,s$^{-1}$.

The He\,I 5876\,{\AA} line becomes more prominent along with the He\,I 6678 and 7066\,{\AA} lines up to at least $+36.3$\,d post peak. These strong He\,I lines are characteristic of typical SNe Ibn. The lines do weaken again at $+70.2$\,d post peak, but still remain the most prominent features. At all times after the initial classification epoch the emission component of these lines dominates over the blue-shifted absorption component.

In the $+70.2$d spectrum the He\,I lines appear to have a noticeable blue asymmetry. This asymmetry was confirmed by comparing fits of the lines in this spectrum to fits of the same lines in the $+36.3$d spectrum. The fits for the He\,I 5876 and 7066\,{\AA} lines are shown in Figure \ref{fig:Line_fits}. The features in the $+36.3$d spectrum are well fit by a single Gaussian with no evident second peak and so we determine such a component is not required. In the $+70.2$d spectrum the same lines are best fit by two Gaussian components, a dominant Gaussian component and an additional, smaller, narrower Gaussian component on the blue side of the dominant Gaussian. In both cases the centre of the dominant Gaussian was fixed for each spectrum. The results of these fits are shown in Table \ref{tab:Spectra}. 

\begin{table*}
    \begin{center}
    \caption{The best fit parameters to the He\,I 5876 {\AA} and He\,I 7066 {\AA} lines of the $+36.3$d and $+70.2$d spectra.}
    \label{tab:Spectra}
    \def\arraystretch{1.5}
    \begin{tabular}{lcccc}
        \hline
        \multirow{2}{*}{Parameter} & \multicolumn{2}{c}{$+36.3$d Spectrum} & \multicolumn{2}{c}{$+70.2$d Spectrum} \\
        \cline{2-5}
        & 5876 {\AA} & 7066 {\AA} & 5876 {\AA} & 7066 {\AA} \\
        \hline
        Primary Peak Centre$^a$ ($10^2$ km\,s$^{-1}$) & \multicolumn{2}{c}{-5.40$^{+0.31}_{-0.32}$} & \multicolumn{2}{c}{-3.90$^{+1.77}_{-1.12}$} \\
        Primary Amplitude (erg\,s$^{-1}$cm$^{-1}$) & 6.58$^{+0.22}_{-0.21}$$\times 10^{-15}$ & 5.30$^{+0.12}_{-0.12}$$\times 10^{-15}$ & 4.55$^{+0.45}_{-0.64}$$\times 10^{-16}$ &  3.81$^{+0.32}_{-0.49}$$\times 10^{-16}$ \\
        Primary FWHM ($10^3$ km\,s$^{-1}$) & 3.40$^{+0.11}_{-0.10}$ & 2.75$^{+0.07}_{-0.07}$ & 2.87$^{+0.21}_{-0.27}$ & 2.17$^{+0.27}_{-0.22}$ \\
        Secondary Peak Centre (km\,s$^{-1}$) & \multicolumn{2}{c}{-} & -1.49$^{+0.06}_{-0.05}$$\times 10^3$ & -1.38$^{+0.15}_{-0.09}$$\times 10^3$ \\
        Secondary Amplitude (erg\,s$^{-1}$cm$^{-1}$) & \multicolumn{2}{c}{-} & 8.32$^{+4.63}_{-2.37}$$\times 10^{-17}$ & 1.39$^{+0.24}_{-0.22}$$\times 10^{-16}$ \\
        Secondary FWHM (km\,s$^{-1}$) & \multicolumn{2}{c}{-} & 8.55$^{+2.75}_{-2.02}$$\times 10^2$ & 1.10$^{+0.36}_{-0.19}$$\times 10^3$ \\
    \end{tabular}
    \end{center}
    $^a$The same peak centre was fit for both spectral features within each spectra.
\end{table*}

\subsection{Polarimetry}

\begin{figure}
    \centering
    \includegraphics[width=0.49\textwidth]{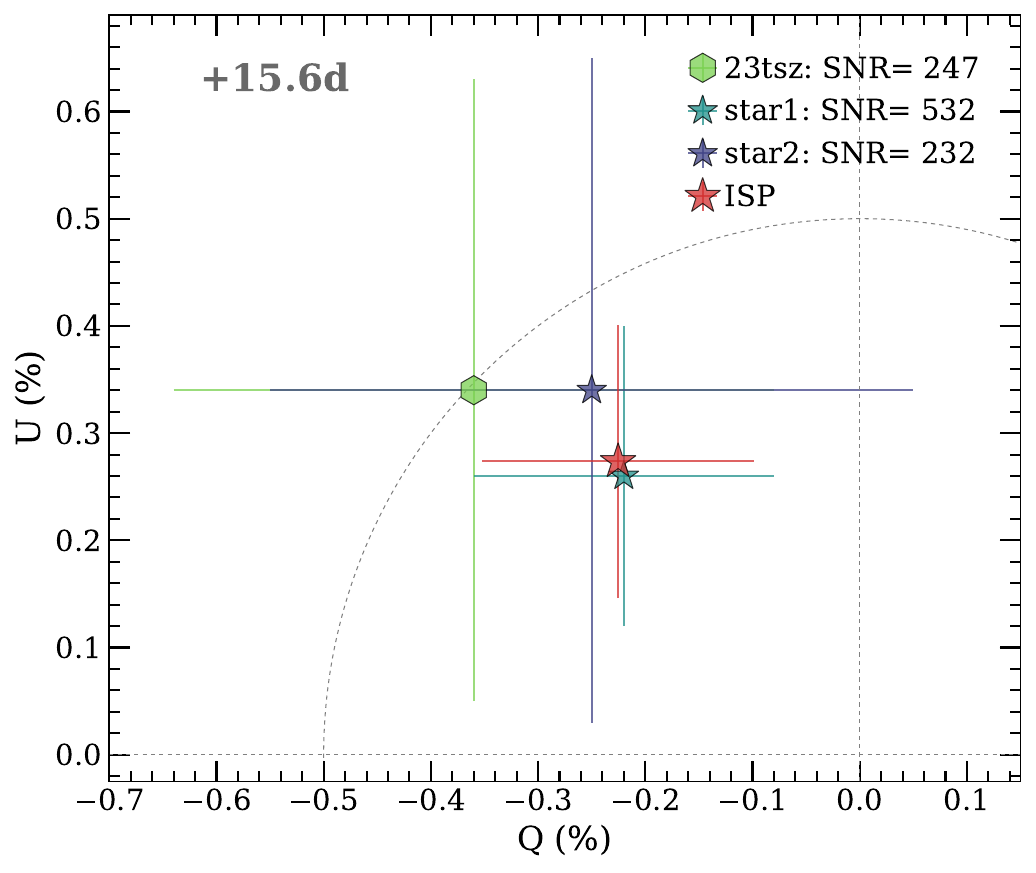}
    \caption{The Stokes $Q$\,--\,$U$ diagram of the $V$-band polarimetry, before ISP correction, taken at $+15.6\,d$. The measured polarization fraction of SN\,2023tsz is consistent with the two other stars present in the image. After correcting for the Galactic ISP and polarisation bias, we find $P=0.08\pm0.31$\%. $Q=0$\% and $P=1$\% have been marked with dotted lines.}
    \label{fig:QU_plane}
\end{figure}

The NOT/ALFOSC V-band polarimetry of SN\,2023tsz taken at $+15.6$\,d appears to be consistent with zero polarisation, as shown in Figure \ref{fig:QU_plane}. For the SN, we find Stokes parameters $Q=-0.36\pm0.28$ and $U=0.34\pm0.29$, but the ALFOSC FOV also covers two nearby bright stars, which are perfectly consistent with the measured polarisation of the SN. Based on the Gaia parallaxes from the third data release  \citep{2023A&A...674A...1G}, the stars are $>150$\,pc above the Milky Way plane, and as such, they probe the full Galactic dust column and can be used to measure the Galactic interstellar polarisation (ISP) component \citep{1995ApJ...440..565T}. We adopt the weighted average of the two stars and find $Q_\mathrm{ISP}=-0.23\pm0.13$ and $U_\mathrm{ISP}=0.27\pm0.13$ for the Galactic ISP. The low value is further supported by the Heiles catalogue \citep{2000AJ....119..923H} which shows 10 stars within 5 $\deg$ from the SN that are consistent with a polarization fraction of $P<0.2$\%. After ISP- and polarisation bias corrections, we find $P=0.08\pm0.31$\% for the SN. 

Whilst we cannot directly estimate the host galaxy ISP, its maximum value should be related to the host galaxy extinction via the following empirical relation: $P_{\mathrm{ISP}} < 9 \times E_\mathrm{B-V}$ \citep{1975ApJ...196..261S}. The photometric and spectroscopic properties of the SN and the host galaxy properties all support the assumption of low host extinction, implying that the host ISP should also be low. As such, we can conclude that the polarisation of SN\,2023tsz is likely intrinsically low. The $V$-band covers mostly continuum and the only clear line feature in the $V$-bandpass is HeI 5876\,{\AA} and that manifests mostly as emission. Line emission is inherently unpolarised, so it can only decrease the observed V-band polarisation, but as the line is narrow ($2.25^{+0.73}_{-0.21}\times10^{3}$\,km\,s$^{-1}$) and covers only a small part of the whole band, the depolarising effect should be negligible. As such, we assume the low polarisation is that of the continuum and conclude that the SN photosphere was consistent with spherical symmetry at $+15.64$\,d post peak.

\subsection{Host data}

\begin{table*}
	\begin{center}
	\caption{Key parameters of different SED fits to the host of SN\,2023tsz, with different SFH burst age parameterisations. The chosen burst age for the best fit (i.e. lowest reduced chi-squared) is given with the model.}
	\label{tab:SED_fits}
	\begin{tabular}{lcccc} 
		\hline
		Model & log(SFR\,[M$_{\odot}$yr$^{-1}]$) & log(Stellar Mass\,[M$_{\odot}]$) & log(sSFR\,[yr$^{-1}]$) & $\chi^2$\\
		\hline
		Scenario 1 (Burst Age=$50$\,Myr) & -2.58 & 7.06 & -9.64 & 0.22\\
		Scenario 2 (Burst Age=$100$\,Myr) & -2.83 & 7.08 & -9.91 & 0.44\\
		Scenario 3 (Burst Age=$5$\,Myr) & -2.95 & 6.64 & -9.59 & 1.74\\
		\hline
	\end{tabular}
    \end{center}
\end{table*}

\begin{figure}
    \centering
    \includegraphics[width=0.48\textwidth]{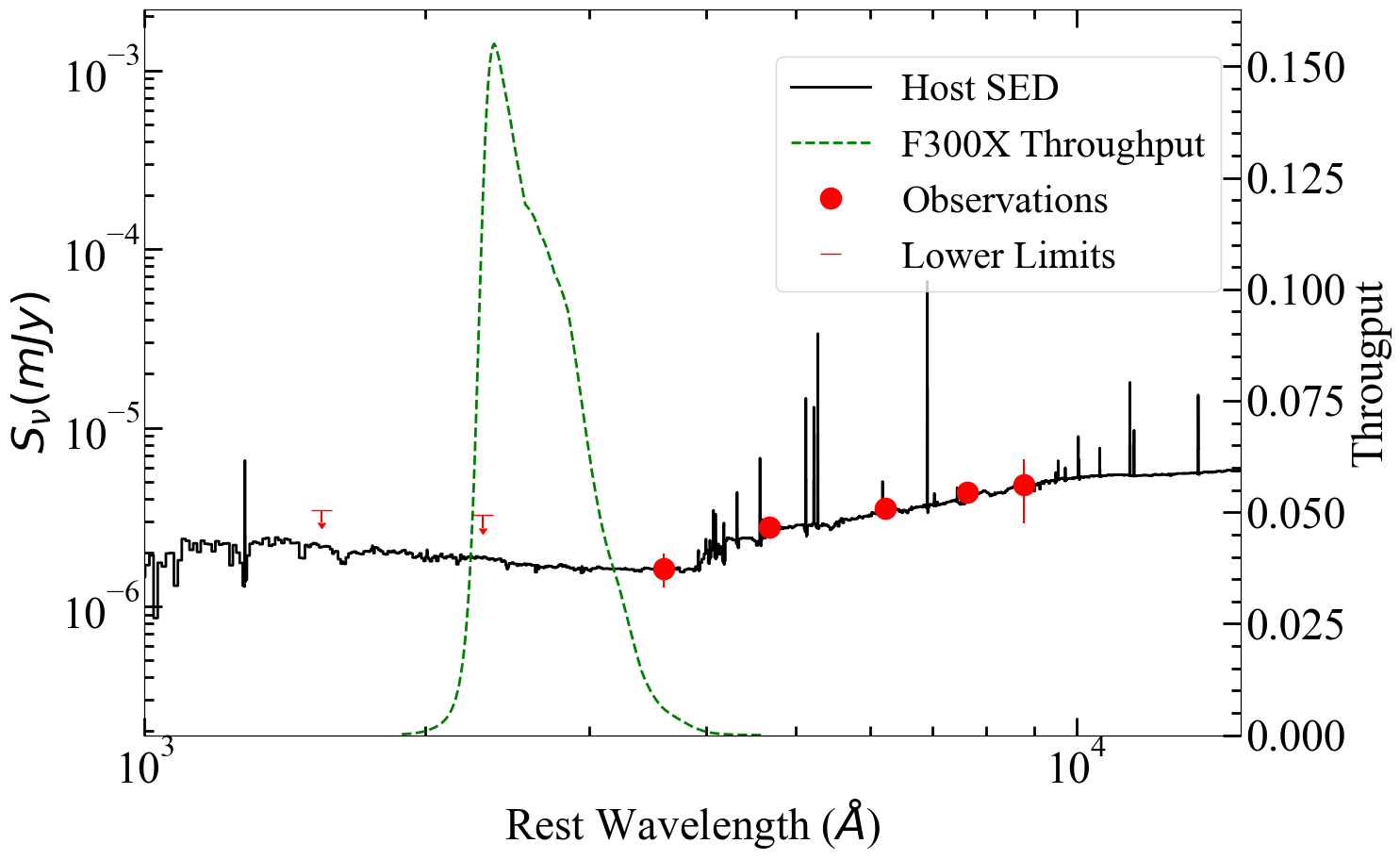}
    \caption{The best fit SED of the host of SN\,2023tsz from CIGALE is shown in black. The observations from KiDs and VIKING are shown by the red points. The green dashed line indicates the region (bandpass and throughput) of the spectrum that would have been observed by the HST F300X filter had the host been at the redshift of PS1-12sk ($z=0.054$).}
    \label{fig:Host_SED}
\end{figure}

\begin{figure*}
    \centering
    \includegraphics[width=\textwidth]{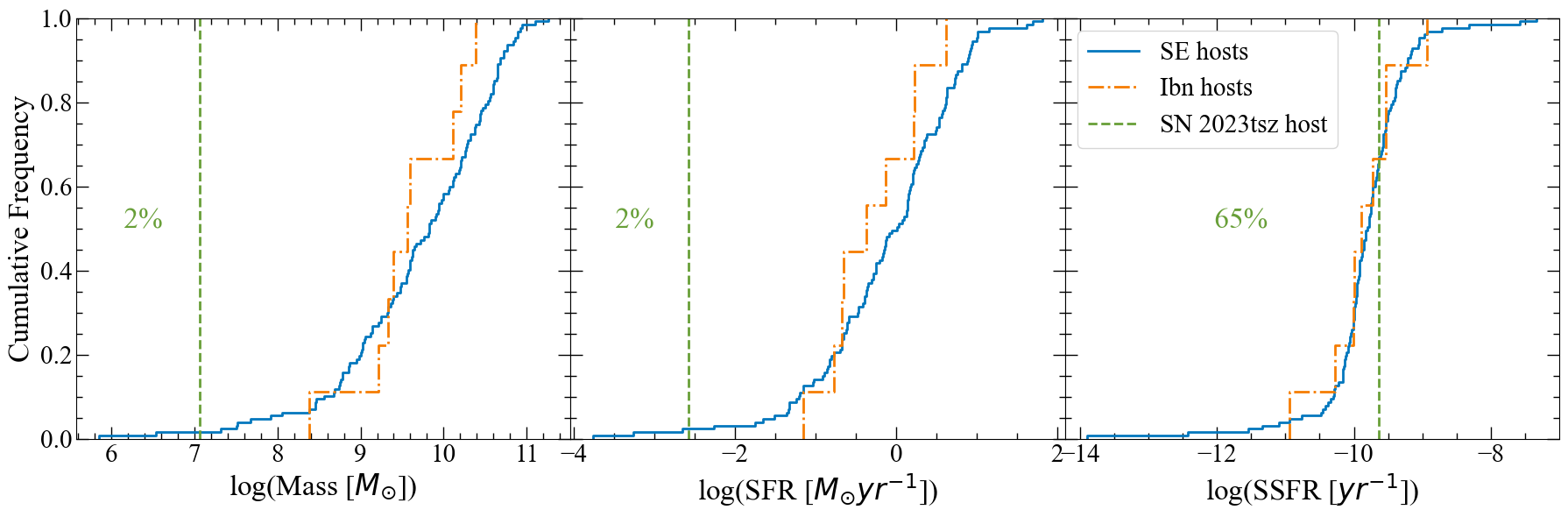}
    \caption{Cumulative frequency plots showing the stellar mass (\textit{left}), SFR (\textit{centre}), and sSFR (\textit{right}) of the host of SN\,2023tsz (dashed green line), SN Ibn hosts (9; dash-dot orange  line) and stripped-envelope SN hosts (9 SN Ibn and 118 SN Ib/c; solid blue). The host properties are from \protect\cite{2021ApJS..255...29S}. The percentage of supernovae hosts with a mass (\textit{left}), SFR (\textit{centre}), or SSFR (\textit{right}) less than the host of SN\,2023tsz is indicated on each plot.}
    \label{fig:CF}
\end{figure*}

We fit the photometric data for the host of SN 2023tsz using the CIGALE software \citep{2019A&A...622A.103B}. CIGALE works by computing composite spectral energy distribution (SED) models of the galaxy using specified modules that each deal with a different component of a galaxy's emission. It then fits these models to the input observations, which for the host of SN\,2023tsz are the survey observations detailed in Table \ref{tab:Surv_log}, and determines the best fitting model, the one with the lowest reduced $\chi^{2}$. Each module has multiple parameters, and the user provides an array of values for each of them. CIGALE then computes a model for each combination of parameter values. Galaxy templates were generated using a delayed star formation history,  using the simple stellar population models of BC03 \citep{2003MNRAS.344.1000B}. Dust attenuation was included using the \cite{2000ApJ...533..682C} law and dust emission following the \cite{2007ApJ...657..810D} templates.

To allow us to compare how the age of the burst affects the other parameters in the SED fitting, three different burst age parameterisations were tested: the first had the age of the burst allowed over a range of ages from 5 to 200\,Myr (scenario 1), the second over a range of ages from 100 to 500\,Myr (scenario 2), and the third from 1 to 5\,Myr (scenario 3). The 5 to 200\,Myr range provided the best fit, with a burst age of 50\,Myr. The best fit model SED, from scenario 1, is shown in Figure \ref{fig:Host_SED}. The key parameters of the best fit in each scenario are shown in Table \ref{tab:SED_fits}. As can be seen, all the fits have comparable values for their SFR, stellar mass, and specific star formation rate (sSFR) meaning the best fit values are robust. For the rest of the analysis, we use the best fit from scenario 1 and its values for SFR, $2.64\times10^{-3}$\,M$_{\odot}yr^{-1}$, stellar mass, $1.15\times10^{7}$\,M$_{\odot}$, and sSFR, $2.29\times10^{-10}$\,yr$^{-1}$. These values are shown in comparison to a sample of stripped envelope (SE) SN host galaxies, from \cite{2021ApJS..255...29S}, in Figure \ref{fig:CF}.

We measure the galaxy to have a radius of $1.6$ arcsec, equivalent to a physical radius of 1.0\,kpc, defined by the radius encompassing 99\% of the galaxy's light in the Legacy Survey DR9 $g$ band images. This equates to an area of $8.2$\,arcsec$^2$ and a physical size of $3.1$\,kpc$^{2}$. This radius is shown in Figure \ref{fig:tsz_loc}. This, along with the SFR from our best fit SED, puts the SFR density of SN 2023tsz's host at $9.3 \times 10^{-4}$ M$_{\odot}$yr$^{-1}$kpc$^{-2}$. When we compare this to the combined data sets of \cite{2019ApJ...871L...9H} and \cite{2018ApJ...855..107G} there are only two SN Ib/c (4.4\% of the sample) and one SN Ibn (5.9\% of the sample) host galaxies with lower SFR densities.

Using the fraction of stars per unit mass that produce CC SNe from \cite{2012MSAIS..19..158B}, $K_{\mathrm{CC}} \sim 0.01$\,M$_{\odot}^{-1}$, the SFR from our best fit SED would mean that we expect a massive star to explode in this galaxy every $\approx$ 35 kyr. As SNe Ibn account for $\approx 1\%$ of CC SNe \citep{2022ApJ...927...25M}, we would expect to observe an event such as 2023tsz in its host approximately once in every $3.5$ Myrs. This expectation could still be too frequent given that lower mass star forming regions, as expected in such a low star-formation host, may under-produce the most massive stars owing to stochastic IMF sampling \citep{2023MNRAS.522.4430S}. 

The only other SN Ibn that was discovered in a low SFR region is PS1-12sk \citep{2013ApJ...769...39S}. No host was found at PS1-12sk's location in HST observations \citep{2019ApJ...871L...9H}. We simulated an observation of the host of SN\,2023tsz at the redshift of PS1-12sk to investigate if it would have been visible in the HST observations. This simulated observation was done using the \textit{PySynphot} package \citep{2013ascl.soft03023S} into which we input the SED generated using CIGALE. The part of the SED visible in the HST F300X filter at $z = 0.054$, the redshift of PS1-12sk,  is shown in Figure \ref{fig:Host_SED}. We obtained an apparent magnitude of $24.61$ mag for the simulated observation, equating to a surface brightness of $26.90$ mag\,arcsec$^{-2}$. This surface brightness is higher than the $5\sigma$ limit, $27.5$\,mag\,arcsec$^{-1}$ measured by \cite{2019ApJ...871L...9H} meaning a host comparable to the host of SN\,2023tsz would have just been visible in their observations.

\section{Discussion} \label{Disc}

\subsection{Comparison of explosion parameters to other SN Ibn/Icn}

The decline rate and peak magnitude of SN\,2023tsz are comparable to the population of SNe Ibn (Figure \ref{fig:Comparison}). When comparing the parameters from fitting the CSM+Ni model to the bolometric light curve of SN 2023tsz with those of other SNe Ibn \citep{2022ApJ...926..125P} and SNe Icn \citep{2022ApJ...938...73P} using the same models, we find that our ejecta mass $M_\mathrm{ej}$ = $0.59^{+0.69}_{-0.38}$\,$M_{\odot}$ is comparable to their values when considering their associated uncertainties. Our value for the CSM mass $M_\mathrm{{CSM}}$ = $0.05^{+0.05}_{-0.02}$\,$M_{\odot}$ is lower than that found for other SN Ibn fit using the same models with the exception of SN\,2019wep which had a comparable value when considering the associated errors. SN\,2023tsz is on the upper end for decline rates of SN Ibn, so this lower value of $M_\mathrm{{CSM}}$ is expected. The large uncertainties on our value of $M_\mathrm{ej}$ are most likely due to our limited photometric coverage of the rise and peak phases, meaning our model fitting is not well constrained at those epochs. The inferred $^{56}$Ni mass is higher than those found for the other SNe Ibn fit using the same model. We take our value for the $^{56}$Ni produced as an upper limit, as our final photometric observation lies below our model fit which means it is consistent with the lower $^{56}$Ni derived for other SNe Ibn. We were unable to obtain later photometry to help further constrain the $M_\mathrm{{Ni}}$ due to SN\,2023tsz becoming too faint for our instruments. In summary, the key parameters (Ni, CSM, and ejecta masses) of SN 2023tsz show general agreement with those of other SNe Ibn when considering associated uncertainties.

\subsection{Spherical Symmetry in SNe Ib/cn}

The polarimetry of SN\,2023tsz is an important addition to the sample of 3 other SNe Ibn/Icn with polarimetric observations. The polarisation factor of SN\,2023tsz, $P=0.08\pm0.31$\%, is consistent with it having spherically symmetric photosphere at $+15.64$\,d post peak. This observed spherically symmetric photosphere does not rule out a progenitor binary system,  as at the time of observation the photosphere will have engulfed the binary system. To date, there are only three other SNe Ibn/Icn with conclusive polarimetry. SN Ibn 2023emq \citep{2023ApJ...959L..10P} and SN Icn 2021csp \citep{2022ApJ...927..180P} exhibited low polarisation, while SN Ibn 2015G showed polarisation up to $\approx2.7$\%, but the exact value is unclear due to an uncertain, but possibly substantial, ISP contribution \citep{2017MNRAS.471.4381S}. While the sample is still small, most SNe Ibn/Icn appear to be consistent with a high degree of spherical symmetry, but at most one appears to have possibly exhibited an aspherical photosphere. Further polarimetry of SNe Ibn/Icn will be needed to either confirm whether this one possible aspherical case is an outlier among a generally spherically symmetric population or if there is population of aspherical interacting SNe.

\subsection{Causes of Asymmetric Line Profiles}

The blue asymmetry seen in SN 2023tsz's $+70.2$\,d post peak spectrum could potentially be explained by a number of scenarios. One scenario is dust formation within the SN ejecta. Dust can partly absorb light from the far side of the ejecta, suppressing the redshifted part of the emission lines. This effect has been observed in some Type II SNe, such as SN 1987A and SN 1999em \citep{10.1007/BFb0114861, 2003MNRAS.338..939E} more than a year after explosion. It typically only appears at later times as the ejecta need to cool below the threshold for dust formation. However, at the epoch of the spectrum we infer a blackbody temperature of T$_{\mathrm{BB}}>9000$\,K which is inconsistent with dust formation in the ejecta. The effect has also been observed in SN\,2006jc at a similar epoch to our observation (\citep{2008MNRAS.389..141M}. \cite{2008MNRAS.389..141M}  had also near-infrared (NIR) observations that showed a second black body component that was consistent with thermal radiation from newly formed dust. As we lack NIR observations we cannot investigate whether this is the case in SN\,2023tsz. Another possibility is that the unshocked CSM is causing the blue asymmetry. If there is CSM in the line of sight it will scatter the photons from the receding ejecta (relative to our line of sight) more than the approaching ejecta, leading to the observed blue-dominant asymmetry. A visual representation of this is given in \cite{2020A&A...638A..92T}. Based on our polarimetric observation, the phtosphere of SN 2023tsz is consistent with spherical symmetry, which implies large fraction of the CSM likely follows a spherical distribution, as expected of a WR star wind. This directly means that there CSM should be in the line of sight, making this scenario a possibility.

\subsection{Implications for SN Ibn Progenitors \& Hosts}

The characterisation study of \cite{2015MNRAS.449.1954P} found that, at the time, all but one Type Ibn SNe were found in spiral galaxies. As they were typically found in star forming regions this supported the idea that their progenitors are massive stars. However, the one exception was PS1-12sk \citep{2013ApJ...769...39S}. As PS1-12sk was found in a region with a low SFR it brought into question whether all SNe Ibn come from massive stars, and if not, then what is the mechanism that strips their progenitors? 

SN\,2023tsz is now the second SN Ibn that has been discovered in a low SFR region comparable to PS1-12sk, although, unlike PS1-12sk, SN\,2023tsz has a visible host. If SN\,2023tsz and PS1-12sk both come from non-massive star progenitors then it could suggest that alternate channels for SNe Ibn are more common than initially thought. However, a low SFR does not rule out a massive star progenitor. The work by \cite{2019ApJ...871L...9H} focused solely on examining the SFR, excluding the sSFR due to the limitations of their HST observations. The sSFR is a better indicator of the current star formation of a galaxy than SFR alone, because sSFR reveals how efficiently the galaxy is forming stars relative to its mass. The sSFR of the host of SN\,2023tsz is in the 65th percentile for the hosts of SE SN, as shown in Figure \ref{fig:CF}. This is not unexpected as lower mass galaxies are observed to have higher sSFR than higher mass galaxies at low redshifts \citep{2013MNRAS.434..209B, 2024arXiv240414499G}. Based on the sSFR alone we cannot rule out a massive star progenitor. This is further supported by lower mass galaxies forming the majority of their stars later than in more massive galaxies \citep{1996AJ....112..839C, 2006A&A...453L..29C, 2019ApJ...874..100T, 2020MNRAS.498.5581B}. 

Estimating the stellar metallicity of the host of SN\,2023tsz using a galaxy mass–metallicity relation (MZR) derived from the observationally supported simulations of \cite{2016MNRAS.456.2140M} gives $\log(Z_{*}/Z_{\odot}) = -1.56$. For single stars, wind driven stripping is weaker at lower metallicity, making it inefficient at producing WRs \citep{2007A&A...473..603M}, although rotation of the star can help in this regard \citep{2005A&A...429..581M}. At these low metallicities binary interactions are thought to be the dominant envelope stripping mechanism \citep[e.g.][]{2001MNRAS.324...33B}, although more recent work suggest that his may not necessarily be the case \citep{2020A&A...634A..79S}. The binary fraction of WR stars appears to be independent of metallicity down to the metallicities of the Magellanic Clouds \citep{2003MNRAS.338..360F}. However, the inferred metallicity of the host of SN\,2023tsz is a factor $\sim$10 lower than that of the Small Magellanic Cloud. Observational constraints on populations at these metallicities from local group dwarfs are extremely difficult. The work of \cite{2003MNRAS.338..360F} is based on only 12 sources, but it does suggest that WR are not ruled out as progenitors in these low-mass hosts, and consequently can not be ruled out as a progenitor for SN\,2023tsz.

Our simulated observation shows that the host of SN\,2023tsz would have just been visible in the observations of the region of PS1-12sk \citep{2019ApJ...871L...9H}. However, we note that SE SNe have been discovered in hosts an order of magnitude less massive \citep{2021ApJS..255...29S} than that of SN\,2023tsz. Such a galaxy would plausibly not have been observable in the observations of \cite{2019ApJ...871L...9H}. We further note the similarity in larger scale environment between SN\,2023tsz and PS1-12sk: in the outskirts of a more massive local galaxy. Although the satellite nature of the host of SN\,2023tsz is uncertain, such a location is to be expected of faint, low-mass satellite galaxies. Given the low luminosity of these galaxies its possible that SNe in such hosts at higher redshifts may be misattributed to being highly offset from a more massive galaxy.

The exceptional nature of the host of SN\,2023tsz then highlights a need for a thorough investigation of the host environments of SNe Ibn. It is not yet clear if they are over-represented in low-mass, low-metallicity galaxies compared to other CCSN types. Any such preference would require explanation by purported progenitor systems. If a population of SNe Ibn exists in low-mass host galaxies, it becomes important to determine whether SNe Ibn from more actively star-forming regions share the same progenitor scenario. Alternatively, the SNe Ibn population may consist of two separate progenitor groups that give rise to SNe with the same observed properties.

\section{Conclusion} \label{Con}
We have presented the analysis of the photometric, spectroscopic, polarimetric, and host properties of SN\,2023tsz, a typical SN Ibn that stands out due to its association with an exceptionally low-mass host galaxy. This study adds to the limited sample of SN Ibn with polarimetric data, making it only the fourth such SN\,Ib/cn to be studied in this way. It is also the first SN Ibn to be discovered, reported, and classified entirely by the GOTO collaboration. The key findings of our analysis are summarized below:
\begin{itemize}
    \item The peak absolute magnitude of SN\,2023tsz, M$_{r} = -19.72 \pm 0.04$\,mag, and its decline rate of $0.145 \pm 0.002$\,mag day$^{-1}$ are consistent with other known SNe Ibn, indicating it is a typical member of the class.
    \item Our modeling of the bolometric light curve using a CSM+Ni model resulted in inferred values of $M_{\mathrm{CSM}} = 0.05^{+0.05}_{-0.02}$\,M$_{\odot}$, $M_{\mathrm{ej}} = 0.59^{+0.69}_{-0.38}$\,M$_{\odot}$, and $M_{\mathrm{Ni}} = 0.04^{+0.01}_{-0.01}$\,M$_{\odot}$, all of which are comparable to other SN Ibn. The high initial decline rate of SN\,2023tsz, compare to other SNe Ibn, is consistent with it having low CSM mass.
    \item At later times, the spectrum of SN\,2023tsz showed blue asymmetry in its prominent emission lines. While the exact cause is uncertain, a plausible explanation involves the scattering of redshifted photons by circumstellar material (CSM), though dust formation in a cool dense shell cannot be ruled out without contemporaneous NIR and mid-infrared data.
    \item Polarimetric observations showed a polarization fraction of $P=0.08 \pm 0.32$\% at 15.6 days post-peak, suggesting a symmetric photosphere. This is consistent with 3 of the 4 polarimetric observations of SNe Ib/Icn to date.
    \item From our SED fitting, the host galaxy of SN\,2023tsz has SFR of $2.64 \times 10^{-3}$\,M$_{\odot}$ yr$^{-1}$, a stellar mass of $1.15 \times 10^{7}$\,M${\odot}$, and a sSFR of $2.29 \times 10^{-10}$\,yr$^{-1}$. While its SFR and stellar mass are in the 2nd percentile for SE SN hosts, its sSFR is in the 65th percentile \citep{2021ApJS..255...29S}.
    \item We estimate a host metallicity of log($Z_{*}/Z_{\odot}$) = $-1.56$ based on the mass-metallicity relation, indicating that the progenitor system of SN\,2023tsz is from an extremely low-metallicity environment.
\end{itemize}

These findings show that SNe Ibn can occur in extraordinarily low-mass, and low-metallicity galaxies. This environment poses problems for the stellar evolution of a single stripped star progenitor, although does not rule them out. An important question that remains surrounding SNe Ibn: Are they over represented in these low-mass galaxies, and if so, why? These host galaxies are often too faint to be detected by contemporary sky surveys, potentially skewing our understanding of the host environments of SNe Ibn. 

The host of SN\,2023tsz shows the need for further investigation into the occurrence rates of SNe Ibn in such hosts. Any over-representation in low-mass, low-metallicity hosts will need to be addressed in future proposed progenitor scenarios.

\section*{Acknowledgements}

BW acknowledges the UKRI's STFC studentship grant funding, project reference ST/X508871/1.

The authors would like to acknowledge the University of Warwick Research Technology Platform (SCRTP) for assistance in the research described in this paper.

This work makes use of observations from the Las Cumbres Observatory global telescope network. Time on the Las Cumbres Observatory network was provided via OPTICON (proposal 23B030).

JL, MP and DO acknowledge support from a UK Research and Innovation Fellowship (MR/T020784/1).

TLK acknowledges support via an Research Council of Finland grant (340613; P.I. R. Kotak), and from the UK Science and Technology Facilities Council (STFC, grant number ST/T506503/1).

For the purpose of open access, the author has applied a Creative Commons Attribution (CC BY) licence to the Author Accepted Manuscript version arising from this submission.

SM was funded by the Research Council of Finland project 350458.

LK and LN thank the UKRI Future Leaders Fellowship for support through the grant MR/T01881X/1.

Based in part on observations made with the Nordic Optical Telescope, owned in collaboration by the University of Turku and Aarhus University, and operated jointly by  Aarhus University, the University of Turku and the University of Oslo, representing Denmark, Finland and Norway, the University of Iceland and Stockholm University at the Observatorio del Roque de los Muchachos, La Palma, Spain, of the Instituto de Astroﬁsica de Canarias. The NOT data presented here were obtained with ALFOSC, which is provided by the Instituto de Astroﬁsica de Andalucia (IAA) under a joint agreement with the University of Copenhagen and NOT.

The Gravitational-wave Optical Transient Observer (GOTO) project acknowledges the support of the Monash-Warwick Alliance; University of Warwick; Monash University; University of Sheffield; University of Leicester; Armagh Observatory \& Planetarium; the National Astronomical Research Institute of Thailand (NARIT); Instituto de Astrofísica de Canarias (IAC); University of Portsmouth; University of Turku, and the UK Science and Technology Facilities Council (STFC, grant numbers ST/T007184/1, ST/T003103/1 and ST/Z000165/1).

DS acknowledges support from the STFC via grant numbers ST/T003103/1, ST/Z000165/1 and ST/X001121/1.

JA's work was funded by ANID, Millennium Science Initiative, ICN12\_009.

TP acknowledges the financial support from the Slovenian Research Agency (grants I0-0033, P1-0031, J1-8136, J1-2460 and Z1-1853).

PC was supported by the Science \& Technology Facilities Council (grants ST/S000550/1 and ST/W001225/1).

RS is funded by a Leverhulme Research Project Grant.

SM was funded by the Research Council of Finland project 350458.

MJD is funded by the UK Science and Technology Facilities Council (STFC) as part of the Gravitational-wave Optical Transient Observer (GOTO) project (grant number ST/V000853/1). 

MRM acknowledges a Warwick Astrophysics prize post-doctoral fellowship made possible thanks to a generous philanthropic donation

AS acknowledges the Warwick Astrophysics PhD prize scholarship made possible thanks to a generous philanthropic donation.

Based on observations collected at the European Organisation for Astronomical Research in the Southern Hemisphere, Chile, as part of ePESSTO+ (the advanced Public ESO Spectroscopic Survey for Transient Objects Survey). ePESSTO+ observations were obtained under ESO program ID 112.25JQ. 

AA acknowledges the Yushan Fellow Program by the Ministry of Education, Taiwan for the financial support (MOE-111-YSFMS-0008-001-P1).

This research has made use of data obtained from the High Energy Astrophysics Science Archive Research Center (HEASARC) and the Leicester Database and Archive Service (LEDAS), provided by NASA's Goddard Space Flight Center and the School of Physics and Astronomy, University of Leicester, UK, respectively.

LG, TEMB, and CPG acknowledge financial support from AGAUR, CSIC, MCIN and AEI 10.13039/501100011033 under projects PID2023-151307NB-I00, PIE 20215AT016, CEX2020-001058-M, FJC2021-047124-I, 2021-BP-00168, and 2021-SGR-01270.

MN is supported by the European Research Council (ERC) under the European Union’s Horizon 2020 research and innovation programme (grant agreement No.~948381) and by UK Space Agency Grant No.~ST/Y000692/1.

TWC acknowledges the Yushan Fellow Program by the Ministry of Education, Taiwan for the financial support (MOE-111-YSFMS-0008-001-P1).

POB was supported by the Science \& Technology Facilities Council (grant number ST/W000857/1).

RK ackowledges support from the Research Council of Finland (grant: 340613).

HK was funded by the Research Council of Finland projects 324504, 328898, and 353019.


\section*{Data Availability}

The underlying raw photometric, spectroscopic, and polarimetric data are available from the relevant data archives. The analysed data are available upon reasonable request to the corresponding author.




\bibliographystyle{mnras}
\bibliography{References} 


\section*{Affiliations}
\small$^{1}$\textit{Department of Physics, University of Warwick, Gibbet Hill Road, Coventry, CV4 7AL, UK.}\\
\small$^{2}$\textit{Institute of Space Sciences (ICE-CSIC), Campus UAB, Carrer de Can Magrans, s/n, E-08193 Barcelona, Spain.}\\
\small$^{3}$\textit{Institut d'Estudis Espacials de Catalunya (IEEC), 08860 Castelldefels (Barcelona), Spain.}\\
\small$^{4}$\textit{Department of Physics \& Astronomy, University of Turku, Vesilinnantie 5, Turku, FI-20014, Finland.}\\
\small$^{5}$\textit{Department of Physics, Lancaster University, Lancaster, LA1 4YB, UK.}\\
\small$^{6}$\textit{European Southern Observatory, Alonso de C\'ordova 3107, Casilla 19, Santiago, Chile.}\\
\small$^{7}$\textit{Millennium Institute of Astrophysics MAS, Nuncio Monsenor Sotero Sanz 100, Off. 104, Providencia, Santiago, Chile.}\\
\small$^{8}$\textit{Graduate Institute of Astronomy, National Central University, 300 Jhongda Road, 32001 Jhongli, Taiwan.}\\
\small$^{9}$\textit{Jodrell Bank Centre for Astrophysics, Department of Physics and Astronomy, The University of Manchester, Manchester, M13 9PL, UK.}\\
\small$^{10}$\textit{Institute of Cosmology and Gravitation, University of Portsmouth, Portsmouth, PO1 3FX, UK.}\\
\small$^{11}$\textit{Astrophysics Research Cluster, School of Mathematical and Physical Sciences, University of Sheffield, Sheffield, S3 7RH, UK.}\\
\small$^{12}$\textit{Instituto de Astrof{\'{i}}sica de Canarias, E-38205 La Laguna, Tenerife, Spain.}\\
\small$^{13}$\textit{Department of Particle Physics and Astrophysics, Weizmann Institute of Science, Rehovot 7610001, Israel.}\\
\small$^{14}$\textit{School of Physics \& Astronomy, Monash University, Clayton VIC 3800, Australia.}\\
\small$^{15}$\textit{Institute for Globally Distributed Open Research and Education (IGDORE).}\\
\small$^{16}$\textit{Astronomical Observatory, University of Warsaw, Al. Ujazdowskie 4, 00-478 Warszawa, Poland.}\\
\small$^{17}$\textit{Cardiff Hub for Astrophysics Research and Technology, School of Physics \& Astronomy, Cardiff University, Queens Buildings, The Parade, Cardiff, CF24 3AA, UK.}\\
\small$^{18}$\textit{Finnish Centre for Astronomy with ESO (FINCA), FI-20014 University of Turku, Finland.}\\
\small$^{19}$\textit{School of Sciences, European University Cyprus, Diogenes Street, Engomi, 1516, Nicosia, Cyprus.}\\
\small$^{20}$\textit{Astrophysics Research Centre, School of Mathematics and Physics, Queens University Belfast, Belfast BT7 1NN, UK.}\\
\small$^{21}$\textit{National Astronomical Research Institute of Thailand, 260 Moo 4, T. Donkaew, A. Maerim, Chiangmai, 50180 Thailand.}\\
\small$^{22}$\textit{School of Physics \& Astronomy, University of Leicester, University Road, Leicester LE1 7RH, UK.}\\
\small$^{23}$\textit{Center for Astrophysics and Cosmology, University of Nova Gorica, Vipavska 11c, 5270 Ajdov\v{s}\v{c}ina, Slovenia.}\\
\small$^{24}$\textit{Instituto de Alta Investigación, Universidad de Tarapacá, Casilla 7D, Arica, Chile.}\\
\small$^{25}$\textit{Dipartimento di Fisica “Ettore Pancini”, Università di Napoli Federico II, Via Cinthia 9, 80126 Naples, Italy.} \\
\small$^{26}$\textit{INAF - Osservatorio Astronomico di Capodimonte, Via Moiariello 16, I-80131 Naples, Italy.}\\
\small$^{27}$\textit{Armagh Observatory \& Planetarium, College Hill, Armagh, BT61 9DG.}\\
\small$^{28}$\textit{Indian Institute of Astrophysics, II Block, Koramangala, Bengaluru-560034, Karnataka, India.}\\
\small$^{29}$\textit{Hiroshima Astrophysical Science Centre, Hiroshima University, 1-3-1 Kagamiyama, Higashi-Hiroshima, Hiroshima 739-8526, Japan.}\\
\small$^{30}$\textit{The Oskar Klein Centre, Department of Astronomy, Stockholm University, AlbaNova, SE-10691, Stockholm, Sweden.}\\
\small$^{31}$\textit{Pondicherry University, R.V. Nagar, Kalapet, Pondicherry-605014, UT of Puducherry, India.}\\

\appendix

\section{Tables}

\begin{table*}
    \renewcommand{\arraystretch}{1.5}
	\centering
	\caption{Spectroscopic time series of SN\,2023tsz.}
	\label{tab:Spec_log}
	\begin{tabular}{cccccccc} 
		\hline
        Date & MJD & Phase (d) & Telescope & Instrument & Grism & R($\lambda/\Delta\lambda$) & Range \AA \\
        \hline
        2023-09-28 & 60216.23 & +4.2 & NOT & ALFOSC & Gr4 & 360 & 3200 - 9600\\
        2023-10-20 & 60237.95 & +25.9 & HCT & HFOSC & Gr7/Gr8 & 500 & 3800 - 9250\\
        2023-10-30 & 60248.28 & +36.3 & NTT & EFOSC2 & Gr13 & 355 & 3685 - 9315\\
        2023-12-03 & 60282.19 & +70.2 & GTC & OSIRIS+ & R1000B+R1000R & 1018, 1122 & 3630 - 100000\\
		\hline
	\end{tabular}
\end{table*}

\begin{table*}
    \renewcommand{\arraystretch}{1.5}
	\centering
	\caption{Survey data of the host of SN\,2023tsz.}
	\label{tab:Surv_log}
	\begin{tabular}{cccc} 
		\hline
        Filter & Mag & Error & Survey \\
        \hline
        $z$ & 22.1993 & 0.3982 & VIKING \\
        $g$ & 22.79302 & 0.04694 & KiDS \\
        $i$ & 22.30737 & 0.06649 & KiDS \\
        $r$ & 22.53202 & 0.03834 & KiDS \\
        $u$ & 23.37166 & 0.20303 & KiDS \\
		\hline
	\end{tabular}
\end{table*}



\bsp	
\label{lastpage}
\end{document}